**Effects of thermal annealing on thermal conductivity of LPCVD silicon carbide thin films**

Lei Tang[a] and Chris Dames[a,b,*]

[a]Mechanical Engineering, University of California, Berkeley, Berkeley, California 94720, USA

[b]Materials Sciences Division, Lawrence Berkeley National Laboratory, Berkeley, California 94720, USA

*Corresponding author: cdames@berkeley.edu

**Abstract**

The thermal conductivity ($k$) of polycrystalline SiC thin films is relevant for thermal management in emerging SiC applications like MEMS and optoelectronic devices. In such films $k$ can be substantially reduced by microstructure features including grain boundaries, thin film surfaces, and porosity, while these microstructural effects can also be manipulated through thermal annealing. Here, we investigate these effects by using microfabricated suspended devices to measure the thermal conductivities of nine LPCVD SiC films of varying thickness (from 120 – 300 nm) and annealing conditions (as-grown and annealed at 950 °C and 1100 °C for 2 hours, and in one case 17 hours). Fourier-transform infrared spectroscopy (FTIR) and X-ray diffraction (XRD) spectra and density measurements are also used to characterize the effects of the annealing on the microstructure of selected samples. Compared to as-deposited films, annealing at 1100 °C typically increases the estimated grain size from 5.5 nm to 6.6 nm while decreasing the porosity from around 6.5% to practically fully dense. This corresponds to a 34% increase in the measured thin film thermal conductivity near room temperature, from 5.8 W/m-K to 7.8 W/m-K. These thermal conductivity measurements show good agreement of better than 3% with fits using a simple theoretical model based on kinetic theory combined with a Maxwell-Garnett porosity correction. Grain boundary scattering plays the dominant role in reducing the thermal conductivity of these films compared to bulk single-crystal values, while both grain size increase and porosity decrease play important roles in the partial $k$ recovery of the films upon annealing. This work demonstrates the effects of modifying the microstructure and thus the thermal conductivity of SiC thin films by thermal annealing.

# 1. Introduction

Silicon carbide (SiC) is important for next-generation microelectromechanical (MEMS) systems[1, 2] and optical and electronic devices[3-6] due to its favorable properties such as large bandgap, high mechanical strength, and high thermal conductivity. The thermal conductivity ($k$) of SiC is important since it determines the amount of heat dissipation that can be managed and thus will affect the final power limits of the devices.

While the $k$ of bulk SiC[7-10] can exceed 400 W/m-K for high-quality single crystals[7] near room temperature, reported measurements of $k$ in nanowires[11, 12] and thin films[13-16] indicate that microstructural effects can reduce $k$ by one to two orders of magnitude. Among these, thin films have continued to attract particular interest due to their application relevance and unique thermal behaviors compared to those of their bulk counterparts[17, 18]. In one theme of work, conventional SiC thin films were grown by sputtering[13, 14] or plasma-enhanced chemical vapor deposition (PECVD)[15], resulting in amorphous and thus low-$k$ (typically 1 - 2 W/m-K) structures due to the limited growth temperatures (typically less than 500 °C). In order to obtain crystalline SiC films, low-pressure chemical vapor deposition (LPCVD) can be used since it allows higher growth temperatures. In addition, LPCVD is known being able to grow better quality thin films and is more suitable for large-scale production than PECVD[19, 20]. Lei and Mehregany[16] used high temperature (~ 900 °C) LPCVD to grow polycrystalline SiC films of thickness 0.93 μm, and reported a high value of $k$ ~ 64 W/m-K near room temperature, though this was measured in a rough vacuum environment (~ 0.1 torr) for which air convection losses may not have been negligible and this would tend to make the reported value of the film $k$ an overestimate. As the dimensions of devices keep shrinking[21], the thermal conductivity of thinner polycrystalline SiC films is important yet unexplored. In addition, none of the previous experiments on thin films[13-16] studied the effects of post-deposition treatments like thermal annealing, which can modify a film's microstructure and thus can also impact the $k$ of SiC thin films.

Therefore, in this paper we use LPCVD to grow a series of polycrystalline SiC films with thicknesses ranging from 120 - 300 nm, and measure their thermal conductivities using a well-established method based on microfabricated suspended devices[22, 23]. In addition, in order to understand the thermal annealing effects on the thermal conductivity, the films are either non-annealed or annealed at two different temperatures (950 °C and 1100 °C) and times (2 hours and 17 hours). For selected samples the effects of the annealing on the microstructure are also analyzed using X-ray diffraction (XRD), Fourier-transform infrared spectroscopy (FTIR), and density determination. Typical grain sizes and porosities are found to range from 5.5 – 6.6 nm and from 6.5% to practically fully dense, respectively. All the thermal conductivity measurements behave in a manner consistent with expectations resulting from the analysis of the grain size and porosity effects and show good agreement with theoretical results calculated from a simple kinetic theory model combined with a Maxwell-Garnett porosity correction. It is found that the near room-temperature thermal conductivity is enhanced by 34% (from 5.8 W/m-K to 7.8 W/m-K) compared to unannealed films when annealing at 1100 °C for at least 2 hours, owing to both grain coarsening and porosity reduction by the annealing.

## 2. Experimental details

### 2.1. Sample preparation

The nine thin-film SiC samples summarized in Table 1 were prepared by the following steps. First, Si substrates were cleaned by submerging them subsequently into a heated Piranha bath at 120 ºC, room temperature dilute HCl (ratio of water to HCl, 1000:1), and room temperature dilute HF (ratio of water to HF, 25:1) for 10 minutes, 2 minutes, and 1 minute, respectively. Then polycrystalline SiC films with two different as-grown thicknesses of 145 nm and 300 nm were deposited on the Si substrates in a commercial LPCVD furnace using dichlorosilane (DCS) and methylsilane dual precursors and $H_2$ as the dilution gas. The flow rates of DCS, methylsilane, and $H_2$ were set at 3 sccm, 30 sccm, and 240 sccm, respectively[24]. In addition, the growth temperature and pressure were set at 800 ºC and 170 mTorr.[24] Only the growth time was varied to obtain different film thicknesses. After completing the deposition processes, some of the films were immediately annealed in an inert nitrogen atmosphere at either 950 or 1100 ºC for 2 hours. One sample (#7) was annealed for a much longer time of 17 hours. Note that the samples' thicknesses were reduced by up to 17% after annealing because they became denser (see Section 3.1.3 for more details). Finally, the nine SiC thermal test devices were created by patterning both the annealed and as-deposited films to the desired dimensions using standard microfabrication techniques (see Supplementary Material Figure S1 and Section S1 for detailed fabrication of SiC devices), and their processing conditions and dimensions are summarized in Table 1.

Table 1. The dimensions (thickness, $t$, and width, $w$) and processing conditions for the samples in this work. All samples have the same length of $L = 18 \pm 0.1$ μm. N/A = not applicable.

| | Sample Dimensions & Processing Conditions | | | | Main Results by Figure Number | | | |
|---|---|---|---|---|---|---|---|---|
| Sample # | Thickness $t$ (nm) | Width $w$ (μm) | Anneal temp. (ºC) | Anneal time (hours) | Effect of sample width (Fig. 3b) | Effect of anneal time (Fig. 4) | Effect of anneal temp. (Fig. 5) | Effect of grain size and porosity (Fig. 6) |
| 1 | 300 ± 3 | 1.10 ± 0.01 | N/A | N/A | | | X | X |
| 2 | 278 ± 3 | 1.10 ± 0.01 | 950 | 2 | | | X | X |
| 3 | 257 ± 3 | 1.10 ± 0.01 | 1100 | 2 | | | X | X |
| 4 | 145 ± 2 | 1.10 ± 0.01 | N/A | N/A | X | X | X | X |
| 5 | 130 ± 2 | 1.10 ± 0.01 | 950 | 2 | | | X | X |
| 6 | 120 ± 2 | 1.10 ± 0.01 | 1100 | 2 | | X | X | X |
| 7 | 120 ± 2 | 1.10 ± 0.01 | 1100 | 17 | | X | | X |
| 8 | 145 ± 2 | 0.60 ± 0.01 | N/A | N/A | X | | | X |
| 9 | 145 ± 2 | 2.10 ± 0.02 | N/A | N/A | X | | | X |

## 2.2. Thermal conductivity measurements

The thermal conductivity measurements were performed using a well-established method based on a microfabricated suspended device[22, 23] as shown in Figure 1(a). The device is built around two SiNx islands ($t_{SiNx}$ ~ 500 nm, and area 30 by 32 μm$^2$). The two islands are denoted as the heating island and the cold island, respectively. Each island has its own serpentine platinum resistance thermometer (PRT) and is supported by six 300 μm-long and 2 μm-wide SiNx legs with platinum lines. Each such island device also incorporates a single SiC thin film ribbon sample that is patterned to bridge the two islands in the device. We designate the dimensions of each SiC ribbon sample as length $L$, width $w$, and thickness $t$, given in Table 1 for all devices.

Figures 1(b) and (c) show optical images of one of the fabricated devices. In addition to the core features just described, each island also incorporates an additional SiC "wing" or "side shelf" of dimensions 30 μm by 10 μm, located on the side of each island opposite the SiC $k$ test sample (for example, the two larger pink rectangles in Fig. 1c). These SiC wings are included as they resulted in a higher yield of successfully released devices after KOH wet etch (step 12 in Figure S1), which we hypothesized was due to the additional mechanical stiffness, and they are fabricated simply by patterning the same deposited SiC film as used for the main $k$ test sample. To minimize the thermal contact resistance to the sample region, the SiNx island pads are patterned to slightly overlap and thus cover each contact edge of the SiC pads and sample, with an overlap of ~ 1 μm, as is apparent schematically in Fig. 1(a, bottom). For two materials in atomically intimate contact like the SiNx-SiC deposition here, the order of magnitude thermal contact resistance[25] is expected to be $R_c'' \sim 10^{-8} \frac{m^2 K}{W}$, and thus the series thermal resistance of the two SiNx-SiC contacts each with 1 μm overlap is estimated to be around 3 orders of magnitude less than the conduction resistance of the SiC ribbon sample itself, and thus is safely negligible. Note that a total of nine similar devices were fabricated, one for each of the SiC film samples listed in Table 1. More details about the device fabrication can be found in Supplementary Material Section S1 and Figure S1.

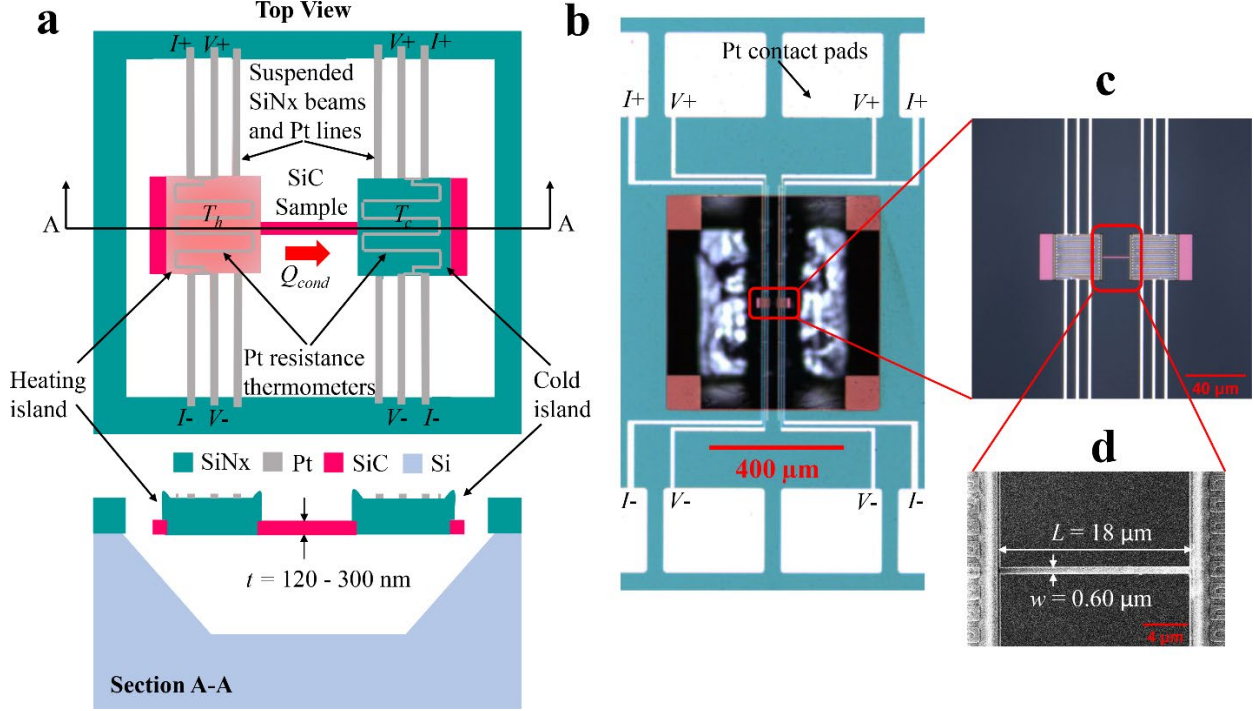

Fig. 1. (a) The top and cross-section schematics of the suspended device used for thermal conductivity measurements of suspended SiC nanoribbons. During the experiments Joule heating is applied to the left island, and the temperature of each island is measured by using four-point resistance thermometry of their respective PRTs. The optical images (b) and (c) show one of the fabricated devices. (d) SEM image showing the dimensions of sample #8.

For measuring the thermal conductivities of the samples, we closely followed Ref.[23]. Briefly, the measurements were accomplished by applying a DC current, $I_{dc}$, to the heating island to create Joule heat and thus a temperature gradient across the SiC ribbon, from the heating island to the cold island. The DC current was swept between -15 µA and 15 µA. At each $I_{dc}$, the 4-probe electrical resistances of the heating and cold islands' respective PRTs, $R_h(I_{dc})$ and $R_c(I_{dc})$, were respectively measured by passing a much smaller AC current of amplitude $I_{ac}$ =1 µA with frequency of 787.7 Hz and 656.6 Hz to each PRT; this $I_{ac}$ is sufficiently small as to cause negligible additional self-heating. The temperature within each island is approximately constant in space ("thermally lumped") due to the fact that the internal thermal resistances of the islands are much smaller than the combined thermal resistance of the sample and supporting beams. The temperature rises on the heating and cold islands, $\Delta T_h(I_{dc})$ and $\Delta T_c(I_{dc})$, with respect to the sample stage temperature (cryostat cold finger temperature ranging from 100 K to 400 K, stabilized with feedback control), were thus measured with each PRT by knowing the $\Delta R_h(I_{dc})$ and $\Delta R_c(I_{dc})$ and their temperature coefficients of resistance (TCR), which were calibrated separately for every device and island (i.e., a total of 18 independent TCRs). The TCRs of the PRTs used in this work are typically in the range of $1 \times 10^{-3}$ - $6 \times 10^{-3}$ K$^{-1}$, similar to those reported in the previous work[23]. Note that each device was measured in a high-vacuum cryostat with a vacuum level of better than 5 x 10⁻⁶ torr; furthermore we surround all sides of the sample with a copper radiation shield that is thermally well connected to the sample stage temperature (i.e., the cold finger of the cryostat), so that convection and radiation losses from the sample can be neglected.

By using a simple energy balance analysis, the thermal conductivity of the SiC sample can be calculated from[23]

$$k = \frac{L}{wt\left(\frac{\partial \Delta T_h/\partial Q}{\partial \Delta T_c/\partial Q}+1\right)\left(\frac{\partial \Delta T_h}{\partial Q}-\frac{\partial \Delta T_c}{\partial Q}\right)} \quad (1)$$

where $Q$ represents the sum of the Joule heat generated from the PRT on the heating island and one current-carrying Pt lead. More details regarding this measurement procedure are explained in Ref.[23].

## 3. Results and discussion

### 3.1. Sample characterization

The samples were characterized using several different techniques. The dimensions of all samples are given in Table 1. All samples have the same length of $L = 18 \pm 0.1$ µm and thicknesses $t$ ranging from $120 \pm 2$ to $300 \pm 3$ nm. The widths and lengths were measured using a Zeiss scanning electron microscope (SEM) with uncertainty better than 2%. An example is shown in Fig. 1(d) for sample #8 with $w = 0.60 \pm 0.01$ µm and $L = 18 \pm 0.1$ µm. The thickness $t$ of each sample with an estimated 2% uncertainty was measured via a spectroscopic ellipsometer (Angstrom Sun Technologies model SE200BM). The density of each SiC film sample, $\rho$, was estimated by measuring the mass difference of the Si substrate before and after the SiC film deposition using an analytical balance (Sartorius model ED124S, with estimated uncertainty less than 0.3%) and using the known volume of the film (wafer area × film thickness from the ellipsometer, and accounting for the film growth on both sides of the wafer). Such density data was then used to estimate the porosity of each sample ($\sigma$) by comparing against the pure bulk SiC density value ($\rho_{bulk} = 3.25$ g/cm³) from Ref.[26]. A Bruker Tensor II Fourier transform infrared (FTIR) spectrometer and Siemens D5000 X-ray diffractometer (XRD) were used to characterize the crystallinity and grain size of the samples.

The FTIR and XRD spectra and the density measurements together are used to analyze the effects of the thermal annealing on the microstructures and thus the thermal conductivities of the samples. Here for simplicity we focus on 257 – 300 nm thick SiC films (samples #1 to #3) to discuss such effects. Figure 2 shows the FTIR and XRD spectra of samples #1 to #3. Table 2 summarizes various characterized parameters.

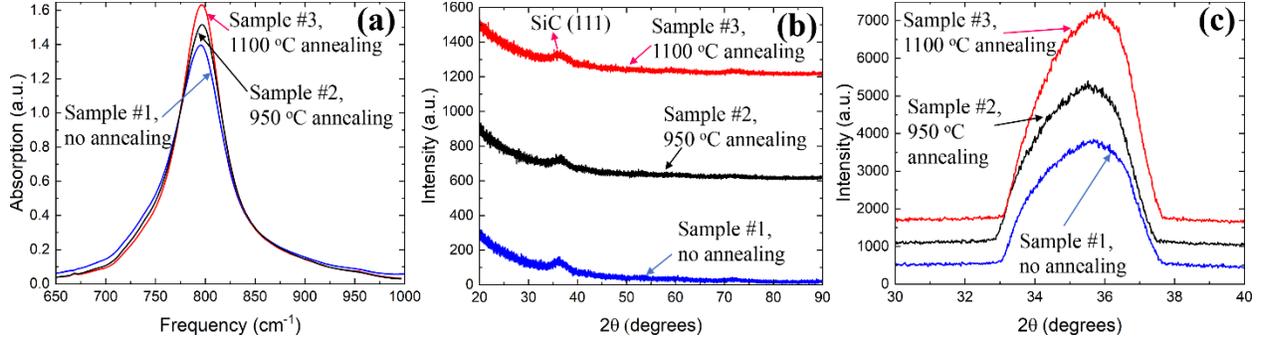

Fig. 2. (a) FTIR spectra, (b) grazing incidence XRD measurements, and (c) XRD rocking curves for Samples #1 - #3. The FTIR and XRD peaks respectively as seen in (a) and (c) become slightly narrower in FWHM after annealing at a higher temperature, as detailed in Table 2. The curves in (b) and (c) have been shifted vertically for easier visualization.

Table 2. The obtained FWHMs of FTIR and XRD rocking curve peaks from Fig. 2(a, c), calculated crystallite sizes ($D_{XRD}$) by applying the Debye-Scherrer equation[27] to the XRD data of Fig. 2c, densities, and nominal porosities, along with the average grain sizes ($D_{fit}$) from fitting the model to $k(T)$ data from Fig. 5a, and the thermal conductivities around 300 K. Annealing slightly densifies the films and coarsens the grains, both of which increase $k$.

| Sample # | FTIR peak FWHM ($cm^{-1}$) | XRD peak FWHM (degrees) | Density, $\rho$ (g/cm$^3$) | Porosity, $\sigma$ (= 1 − $\rho/\rho_{bulk}$), using $\rho_{bulk}$ = 3.25 g/cm$^3$ | $D_{XRD}$ (nm) | $D_{fit}$ (nm) | $k$ @ 300K (W/m-K) |
|---|---|---|---|---|---|---|---|
| 1 (unannealed) | 62 | 3.38 | 3.04 | 6.5% ± 1.0% | 2.7 | 5.5 ± 0.2 | 5.8 ± 0.2 |
| 2 (950 °C) | 58 | 3.32 | 3.19 | 1.8% ± 1.1% | 2.8 | 5.6 ± 0.2 | 6.5 ± 0.2 |
| 3 (1100 °C) | 52 | 3.16 | 3.27 | -0.6% ± 1.2% | 2.9 | 6.6 ± 0.2 | 7.8 ± 0.2 |

### 3.1.1 Effects of annealing on FTIR spectra

Considering first the FTIR spectra, Figure 2(a) shows that all three samples exhibit a peak around 796 cm$^{-1}$, corresponding to the transverse optical phonon mode of β-SiC[28]. The FWHM gradually shrinks as the annealing temperature increases, that is, from sample #1→#2→#3. This indicates that the level of crystallinity of SiC becomes higher as the annealing temperature increases[29].

### 3.1.2 Effects of annealing on XRD spectra

Turning to the XRD spectra, the grazing incidence XRD of Figure 2(b) shows the SiC (111) peak at 2θ of ~ 36 degrees[30]. Thus, we performed the rocking curve scan of XRD around this peak to more carefully determine its FWHM as shown in Fig. 2(c). Similar to the FTIR peak, the breadth of this XRD peak also decreases slightly with increasing annealing temperature, indicating the larger grain size. Based on this XRD data, the average SiC crystallite sizes, $D_{XRD}$, are calculated by using the Debye-Scherrer equation[27], as summarized in Table 2. The obtained shapes of the peaks shown in the rocking curve scans of XRD deviate notably from the ideal Gaussian behavior, which may be due to defects and strain[31, 32]. These potential imperfections of the films along with inherent instrument limitations can also cause XRD peak broadening to a final effective FWHM larger than the ideal Debye-Scherrer value. This may explain why the $D_{XRD}$, which is inversely proportional to the FWHM, is smaller than the $D_{fit}$ obtained from fitting the model to $k(T)$ data (also see Section 3.3).

### *3.1.3 Effects of annealing on density/porosity*

Lastly, Table 2 shows that the estimated density of the SiC films also increases with the annealing treatment. The densest sample is obtained by annealing at the highest temperature of 1100 ºC and has a density of 3.27 g/cm$^3$, which is within 1% of the bulk SiC value[26] of 3.25 g/cm$^3$, while the as-deposited sample has an estimated porosity of 6.5%. The estimated uncertainty of this porosity determination is around ±1.0%, and therefore the apparently non-physical negative $\sigma$ of Sample #3 is within experimental uncertainty of simply being fully dense ($\sigma = 0$), which is how we treat it in the modeling section below.

All the characterization results summarized in Table 2 indicate that higher annealing temperature tends to improve the sample quality with reduced porosity and larger average grain size, both of which tend to increase $k$. This indeed is observed in the room-temperature values of $k$ which are also included as the final column of Table 2, as well as in the fit effective grain size $D_{fit}$ determined from a microscale model for the thermal conductivity that will be explained in Section 3.3.

### *3.2. Thermal conductivity measurement results*

### *3.2.1. Verification of the reliability of the suspended device*

We first assess the repeatability of the measurements by measuring the temperature rises on the heating and cold islands a total of twelve times for each heating power and nominal stage temperature (i.e., cryostat cold finger temperature). These 12 measurements are obtained by sweeping the stage temperature (3 repeats of each $T$ from 150 to 350 K) and $I_{dc}$ (4 repeats of each $I_{dc}$ for each stabilized stage temperature). Specifically, the stage temperature is ramped in 50 K steps from 100 K up to 400 K and back down to 100 K. After stabilizing the sample at each stage temperature, we take one $k$ measurement set by sweeping $I_{dc}$ in 1 µA steps from 0 to -15 µA, back up to 15 µA, and finally back down to 0. As a representative example, Figure 3(a) shows raw data for Sample #9 obtained at the three visits to the stage temperature of 300 K (visits denoted #1, #2, #3 in the plot legend). Focusing on the golden diamond points for visit #3, the measured $Q(I_{dc})$ and $\Delta T_h(I_{dc})$ show outstanding repeatability: for example, around $Q = 1.15$ µW and $\Delta T_h = 4.8$ °C, there are four different gold diamonds plotted (corresponding to the four repeated currents of $I_{dc} = $ [-14, -14, 14, 14] µA), all of which are virtually indistinguishable from each other. Similarly, the $Q(I_{dc})$ and $\Delta T_h(I_{dc})$ data for stage temperature visits #1 and #2 are also virtually indistinguishable (red squares and green circles, respectively). Finally, all of this data is very well fit by the linear model of Eq. (1), as shown by the solid lines in Figure 3(a).

The method is further validated by comparing the measured thermal conductivities of the as-deposited samples with the same thickness of 145 nm but three different widths of 0.60, 1.10, and 2.10 μm (samples #4, #8 and #9 listed in Table 1). Note that these three samples are fabricated from the same SiC film, and thus they should have the same microstructural parameters including the grain sizes and porosities. Figure 3(b) shows their measured thermal conductivities at different temperatures. Here, each point's plotted temperature $T$ is the average of that point's $T_c$ and $T_h$, which themselves are represented respectively as the lower and upper limits of the x-axis error bars (which are smaller than the points themselves). The uncertainties of the thermal conductivity measurements represented as the y-axis error bars stem from the errors when determining the dimensions of the samples and electrical resistances of the PRTs. Note that the error bars shown in the rest of work are obtained in a similar way.

Figure 3b clearly shows that the obtained thermal conductivities of these three samples agree quite well with each other (the worst case discrepancy is 3%), so the width in the range of 0.60 – 2.10 μm has no noticeable impact on the measured $k$. This is expected on physical grounds because the dominant heat carrying phonons are mainly scattered by the pores and grain boundaries (recall from Table 2 that the estimated grain size diameter is only ~ 5 nm), which are the same in all three samples since they were microfabricated from the same SiC film deposition. In contrast, the additional phonon scattering by far-away sample edges at the scale of 0.6 - 2.1 μm is completely negligible by comparison. More details about the thermal conductivity physics and scattering mechanisms will be discussed in Section 3.3.

Furthermore, the excellent consistency among the three samples in Fig. 3b is also an important verification that the measurement scheme in this work is practically free from parasitic effects such as radiation and convection losses, spreading resistances within each island, or other artifacts which in general would scale with sample width in a different way than the pure 1D heat conduction framework that is built into the model of Eq. (1). Further reassurance in this regard is evident from Fig. 5 which shows that the $k$ results are practically independent of the sample thickness as well. Recall also from the fabrication process as discussed in Section 2.2 that the SiNx film of the heating and cold islands is directly grown on the SiC layer and is patterned to slightly cover the edges of the SiC sample bridge itself by around 1 μm, which helped ensure that the contact resistances between the SiNx islands and the SiC ribbon were negligible.

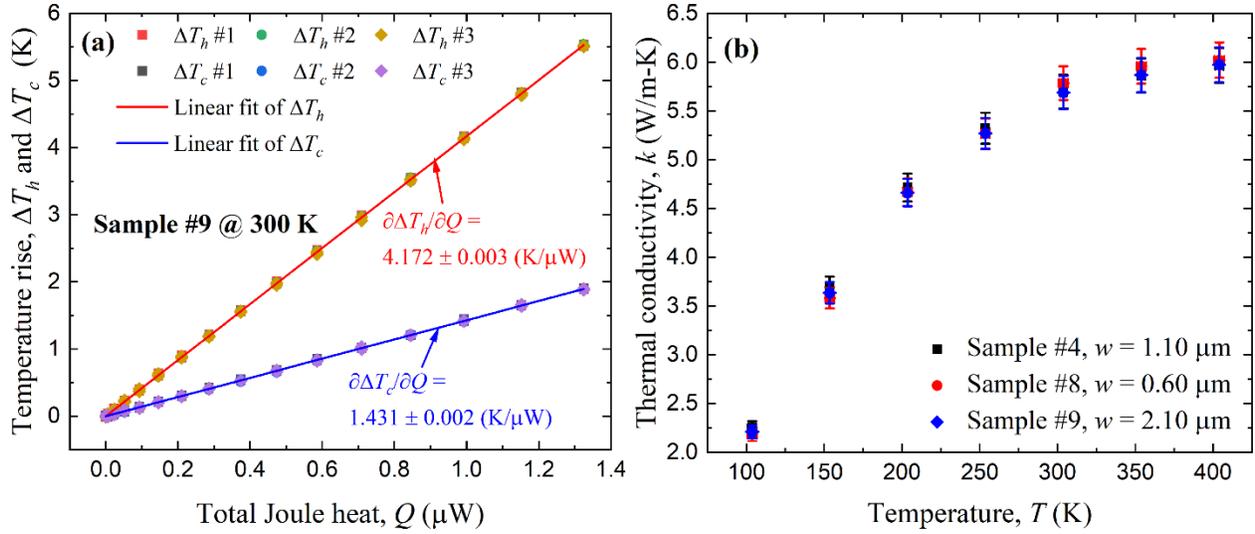

Fig. 3. (a) Measured temperature rises on the heating and cold islands as functions of heating power for sample #9 at a stage temperature of 300 K. The results for three consecutive measurement trials are highly linear and virtually indistinguishable from each other (points overlap), which helps confirm the stability and precision of the experiments. (b) Measured thermal conductivity as a function of temperature for the three samples of $t$ = 145 nm and no annealing, namely #4, #8 and #9. The measured $k$ is highly repeatable across samples (generally to within 1 - 2%), and as expected the width has no noticeable impact on the measured $k$.

### *3.2.2. Effects of annealing time on thermal conductivity*

We next explore the effects of the annealing time on the thermal conductivity of SiC. Figure 4 shows the measured thermal conductivity as a function of temperature for the samples with the same annealing temperature (1100 °C) but with two different annealing times of 2 hours and 17 hours (samples #6 and #7). The results for the as-deposited sample (sample #4; unannealed) are also plotted for comparison. As can be clearly seen, the thermal conductivities of samples #6 and #7 are very close to each other, within 3%, even though #7 has been annealed for a much longer time. Thus, since annealing longer than 2 hours has a negligible impact on the thermal conductivity in Fig. 4, and since phonon scattering in these samples is dominated by microstructural features (porosity and grain boundaries; see also the model in Section 3.3), this strongly suggests that the microstructural evolution during an 1100 °C anneal has largely completed within 2 hrs.

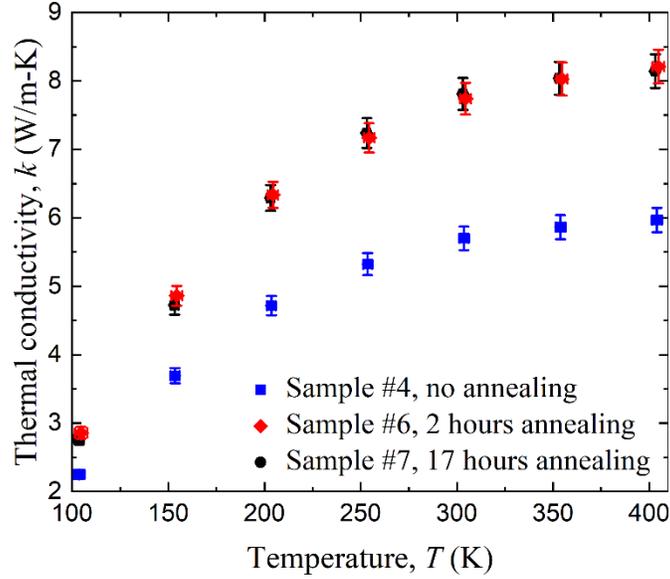

Fig. 4. Effect of annealing time on $k$, for an approximately constant sample cross section of (120 - 145 nm) × 1.10 μm and annealing temperature of 1100 °C. The results for samples #6 and #7 are very similar to each other, which shows that annealing longer than 2 hours has negligible further effect on changing the thermal conductivity.

### *3.2.3. Effects of annealing temperature on thermal conductivity*

Beyond the annealing time, we have also explored the effects of the annealing temperature on the thermal conductivity through samples #1 - #6. As a reminder, the samples #1 - #3 and #4 - #6 respectively have thicknesses of 257 – 300 nm and 120 – 145 nm. In addition, the samples #1 and #4, #2 and #5, and #3 and #6 respectively are non-annealed, annealed at 950 °C for 2 hrs, and at 1100 °C for 2 hours. As can be seen from Fig. 5a (thicker samples) and 5b (thinner samples), $k$ increases with the annealing temperature, while the thickness effect on $k$ is negligible. This is expected and is consistent with the microstructural evolution as summarized in Table 2: with increased annealing temperature the FTIR and XRD peaks become narrower indicating larger average grain sizes, while the porosities also decrease, all of which suggest increased $k$. In addition, comparing panels of Fig. 5 shows that the thermal conductivities between the relatively thicker (a) and relatively thinner (b) films are very close to each other, which confirms that the overall phonon scattering is dominated by microstructural effects rather than film surface scattering. This is also expected because all six of these films were grown using the same growth conditions, with the only difference being the growth time as mentioned in Section 2.1, i.e. we would expect the as-grown films #1 and #4 to have similar microstructures. Overall in these results the thermal conductivity was enhanced by up to 38% when annealing at the highest temperature of 1100 °C considered in this work, as compared to the equivalent as-deposited film.

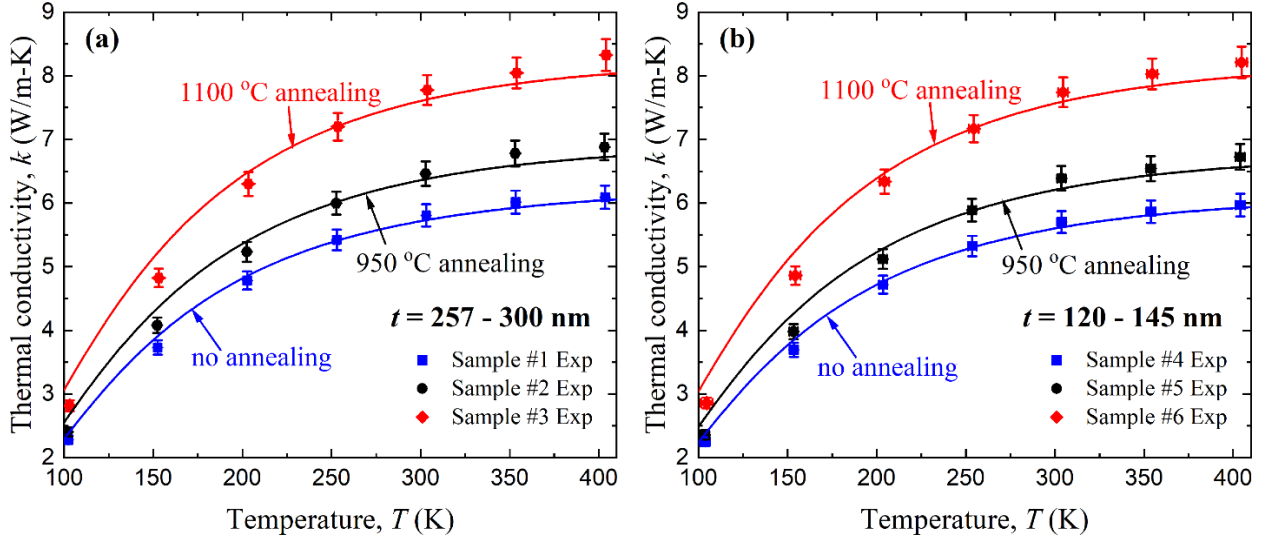

Fig. 5. Effects of annealing temperature at constant time (2 hours duration), for thicker [(a), $t = 257 – 300$ nm] and thinner [(b), $t = 120 – 145$ nm] films all at constant width of 1.10 μm. Experimental results (Exp) are shown by symbols while the best fits obtained from the kinetic theory model combined with the Maxwell-Garnett porosity correction are shown by solid lines.

### 3.3. Theoretical model

#### 3.3.1. Model development

The physics underlying the thermal conductivity changes in these samples is explained by using kinetic theory[17] combined with the Maxwell-Garnett result for spherical pores[33, 34]. First, to account for the effects of porosity we use the classical Maxwell-Garnett result for pores of negligible thermal conductivity,

$$k_{Eff} = k_{Matrix} \cdot \frac{1-\sigma}{1+\frac{1}{2}\sigma} \approx k_{Matrix} \cdot \left(1 - \frac{3}{2}\sigma\right) \qquad (2)$$

where $k_{Eff}$ is the measured thermal conductivity of the actual porous sample (simply denoted as $k$ in all the experimental figures $k(T)$ above, obtained using Eq. (1)), $\sigma$ is the sample porosity estimated from the density measurements as in Table 2, and $k_{Matrix}$ is the intrinsic thermal conductivity of the matrix in the absence of porosity but still including the effects of phonon scattering by grain boundaries, point defects, and phonon-phonon scattering. Recall from Table 2 that Sample #3 had a nominal $\sigma$ that was slightly negative which we attribute to the experimental uncertainty in the $\sigma$ estimation; because we consider that to be non-physical we simply set $\sigma = 0$ for the model of Sample #3. In addition, although the densities and thus the porosities were measured only for Samples #1 - #3 as shown in Table 2, the same porosity value is used in the model if the samples have the same annealing temperature (i.e., assumes that Samples #1, #4, #8 and #9 have the same porosity, likewise Samples #2 and #5, and Samples #3, #6 and #7). We expect this assumption to be reasonable since all the films were initially grown using the same recipe varying only the growth time, and its suitability is confirmed by the fact that the final $k$ values are impacted only by the annealing temperature but not anneal time nor film thickness, as demonstrated in Figs. 4 - 6. The final approximate form of Eq. 2 holds for small $\sigma$, and is a very good approximation in the present work.

Then following Ref.[35] we use kinetic theory to express the phonon thermal conductivity of the matrix as,

$$k_{Matrix} = \frac{1}{3}\Sigma_{pol} \int Cv\Lambda_{eff} d\omega \qquad (3)$$

where $C$ is the volumetric heat capacity, $v$ is the group velocity, $\omega$ is the angular frequency, $\Lambda_{eff}$ is the effective mean free path, and the summation accounts for the three acoustic phonon branches (see Supplementary Material Section S2 for further details).

In general the effective mean free path includes all the intrinsic scattering mechanisms in high-quality bulk single-crystal SiC, $\Lambda_{bulk}$ (principally phonon-phonon scattering at these temperatures), as well as the additional scatterings due to grain boundaries, $\Lambda_{gb}$, defects, $\Lambda_d$, and physical boundaries of the samples, $\Lambda_{bdy}$. The effective mean free path combines all these scatterings in parallel and is calculated using Matthiessen's rule,

$$\Lambda_{eff}^{-1} = \Lambda_{bulk}^{-1} + \Lambda_{bdy}^{-1} + \Lambda_{gb}^{-1} + \Lambda_d^{-1} \qquad (4)$$

A simple model for $\Lambda_{bulk}(\omega,T)$ can be determined independently by fitting bulk SiC $k(T)$ data obtained from Ref.[7] (see Supplementary Material Section S2 for details). For the boundary scattering at the surfaces of our suspended thin film we use a result for a diffuse rectangular rod[36], $\Lambda_{bdy} = \frac{3t}{4}\left(\ln\frac{2w}{t} + \frac{t}{3w} + \frac{1}{2}\right)$. In practice we find this $\Lambda_{bdy} \gg \Lambda_{eff}$ for all our samples and film boundary scattering effects are at most a few percent impact on the final $k$ in this study. For the grain boundary scattering we use a nongray model[35], with $\Lambda_{gb} = C_1/\omega$ where $C_1$ is directly proportional to the average grain size as discussed further below. Finally, for point defect scattering we use a Rayleigh-like scattering model[37, 38], $\Lambda_d = C_2 v/\omega^4$, where $C_2$ is inversely related to the (unknown) concentration of point defects.

Thus this model has two free parameters, $C_1$ and $C_2$, which we determine independently for each sample by fitting its experimental $k(T)$ data. However, we find in practice that for all 9 samples the best fit is insensitive to the point defect scattering parameter once $\frac{C_2}{(2\pi)^4}$ exceeds around $6 \times 10^{42}$ s$^{-3}$, and thus the point defect scattering is unimportant for the thermal conductivity model. See SI section S3 for details. Physically this indicates that any point defect scattering in this model is negligible compared to the other mechanisms, or equivalently, $\Lambda_{eff}^{-1} \gg \Lambda_d^{-1}$.

Accordingly, we simplify the model by dropping the defect scattering term from Eq. (4), reducing the model to a one-parameter fit ($C_1$) for each sample.

Thus the final results for $\Lambda_{eff}$ are dominated by $\Lambda_{gb}$ in the regime of our experiments, i.e., $\Lambda_{eff}^{-1} \approx \Lambda_{gb}^{-1}$ to within a few percent. Six examples of the resulting best fit curves are shown in Fig. 5. Finally we can relate each sample's best-fit $C_1$ value back to an equivalent effective average grain size, $D_{fit}$, using[35],

$$D_{fit} = C_1/\alpha\beta\omega_0, \qquad (5)$$

where $\alpha = 0.72$, $\beta = 0.70$, and $\omega_0 = 7.5 \times 10^{13}$ rad/s is the maximum angular frequency of the model dispersion relation (see Supplementary Material). Three examples of these resulting $D_{fit}$ are given in Table 2 which shows that the $D_{fit}$ are comparable to the $D_{XRD}$ estimated by the Debye-Scherrer expression, and also that $D_{fit}$ increases slightly with the intensity of annealing, a trend that is further confirmed in Supplementary Material Table S2 for the $D_{fit}$ of all 9 samples.

### 3.3.2. Combined effects of grain size and porosity on thermal conductivity

Having established and verified the $k$ model above, we can now use it to better visualize the effects of both grain size and porosity on $k$ over a wider parameter range, which may be helpful for guiding the thermal design of SiC films in various applications[1-6]. See Figure 6, which also considers two different film thicknesses. As can be seen, $k$ increases almost linearly with increasing $D_{fit}$ and/or decreasing $\sigma$, a functional dependency that can be inferred from Eqs. (2)-(5) which yield $k \propto D_{fit} \cdot \left(1 - \frac{3}{2}\sigma\right)$ when both $D_{fit}$ and $\sigma$ are small. In addition, while the physical boundary scattering at the film surfaces becomes significant for large grain sizes (as indicated by the differences between solid and dashed lines, representing 300 nm and 145 nm thick films, respectively), the impact of such $\Lambda_{bdy}$ on $k$ is negligible for our samples with their very fine grains as seen in the inset.

The inset summarizes the room-temperature $k$ of all 9 samples. Recall the assumption that was made for the samples' porosities, that is, we assume that the samples have the same porosity if they have the same annealing temperature. Thus, the $k$ measurements for Samples #1, #4, #8 and #9 should be compared to the model with a porosity of 6.5% in Fig. 6 (i.e., blue symbols and blue lines), and similarly compare black symbols to black lines for $\sigma = 1.8\%$, and red symbols to red lines for $\sigma = 0$. In all cases the agreement between model and experiment is excellent. We also find that, compared to single crystal bulk SiC with its $k \approx 490$ W/m-K at room temperature[7], the reduced $k$ in these samples is dominated principally by grain boundary scattering, with porosity also playing a role of up to ~ 10% additional reduction in $k$. In contrast, scattering by thin film surfaces, point defects, and phonon-phonon scattering are all negligible in these samples (representing ~ 1% or less of the thermal resistance). Lastly, the room-temperature $k$ in these samples increases from 5.8 to 7.8 W/m-K by increasing the annealing temperature up to 1100 °C since the grain sizes increase from 5.5 to 6.6 nm while the porosities decrease from 6.5% to practically fully dense. The obtained $D_{fit}$ grain sizes for this $k$ model are comparable to and follow the same trends as those extracted from the FWHMs of the XRD peaks as seen in Table 2. As such, we conclude that the observed thermal conductivity enhancement upon annealing in this set of samples is mainly due to the increased grain sizes and decreased porosities.

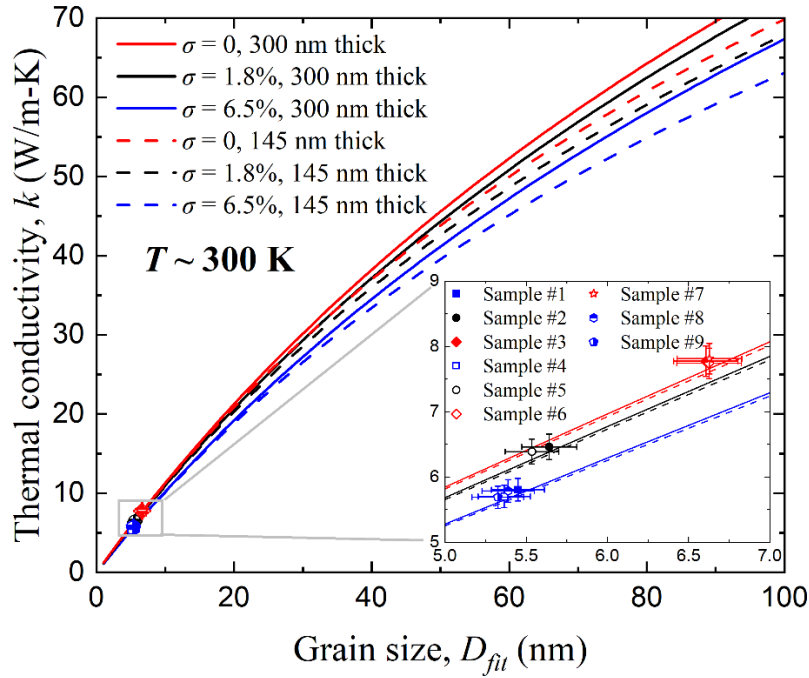

Fig. 6. Combined effects of grain size and porosity on the thermal conductivity. The lines are calculated at $T = 300$ K using the kinetic theory model combined with the Maxwell-Garnett porosity correction. The experimental results (points, from $T$ close to 300 K) agree well with the theoretical model, and show $k$ increases with increasing grain size and decreasing porosity. This figure shows that $D_{fit}$ and $\sigma$ (but not film thickness or width) are the dominant mechanisms by which the sample preparation details impact $k$.

## 4. Conclusions

In summary, we used microfabricated suspended devices to measure the thermal conductivities of nine LPCVD SiC thin films with thicknesses ranging from 120 – 300 nm. We also analyzed the effects of thermal annealing on both the microstructures and the thermal conductivities. Specifically, the SiC samples were either non-annealed or annealed at two different temperatures (950 ºC and 1100 ºC) and times (2 hours and 17 hours). The samples were also characterized using FTIR, XRD, and an estimate of density. When the annealing temperature increased to 1100 ºC, the grain size typically increased from 5.5 to 6.6 nm and the porosity decreased from around 6.5% to essentially fully dense. The measured thermal conductivity at ~ 300 K was enhanced by 34% when annealing at 1100 °C for at least 2 hours as compared to unannealed films. Such enhancement is expected as it is consistent with the trends in the effects of annealing on the grain size and porosity. It may be possible to further enhance $k$ by annealing at an even higher temperature, although it is limited to 1100 °C in this work. All the measurements can be well fit by a simple model combining phonon kinetic theory and grain-boundary scattering with a Maxwell-Garnett porosity correction (agreement between fit and measured $k(T)$ is within 3%). The results shown in this work reveal the ability of modifying the microstructure and thus $k$ of SiC thin film by annealing, which is relevant for thermal management in potential SiC applications ranging from MEMS to optoelectronics.

## Supplementary Material

See the *supplementary material* for additional information on the sample and device fabrication, the mean free paths of bulk SiC and thermal conductivity calculations, the fit average grain sizes of all 9 samples, and the example of the fit using two free parameters.

## Acknowledgements

Financial support from the Marjorie Jackson Endowed Fellowship Fund, the Howard Penn Brown Chair, and the National Science Foundation (ENG-CBET, Award #2020842) is gratefully acknowledged. This work was partially performed at the UC Berkeley Marvell Nanolab. The authors appreciate the support of the staff and facilities that made this work possible.

**Supplementary Material**

**Effects of thermal annealing on thermal conductivity of LPCVD silicon carbide thin films**

Lei Tang[a] and Chris Dames [a,b,*]

[a]Mechanical Engineering, University of California, Berkeley, Berkeley, California 94720, USA

[b]Materials Sciences Division, Lawrence Berkeley National Laboratory, Berkeley, California 94720, USA

*cdames@berkeley.edu

# S1. Sample and device fabrication

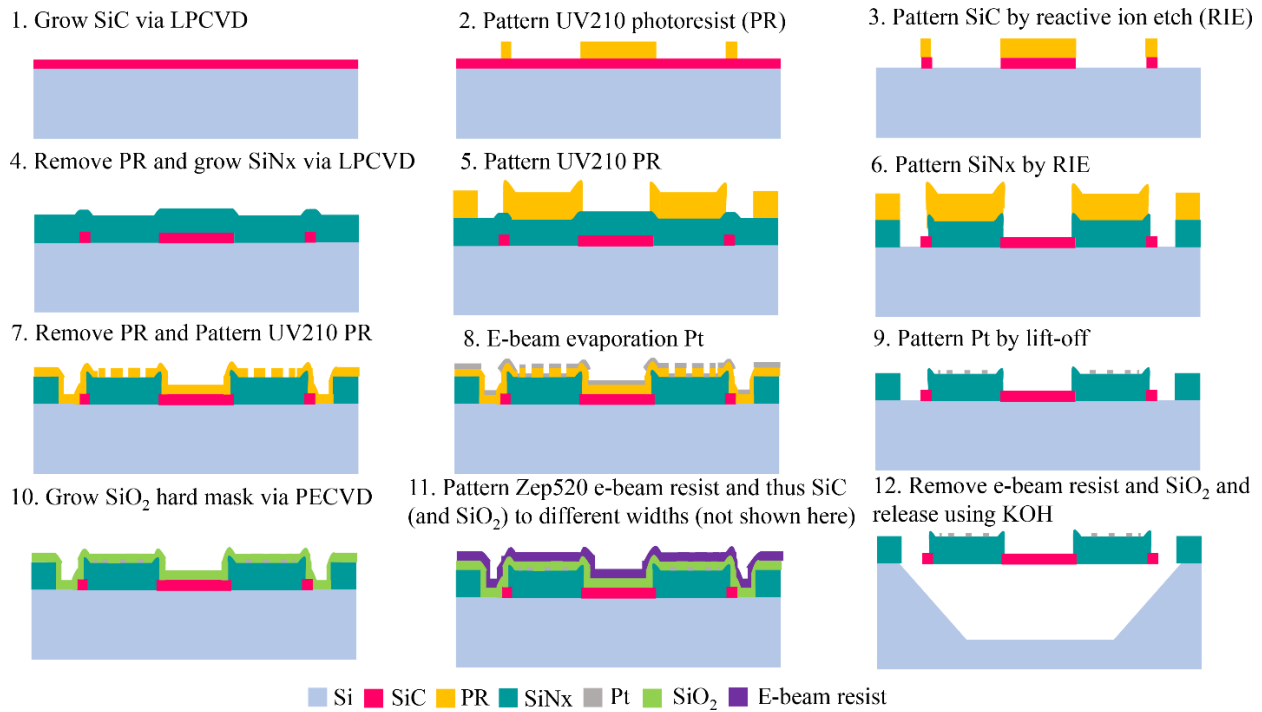

Fig. S1. Device and sample fabrication process. Step 11 patterns the SiC ribbon to its final width in the dimension out of the page (denoted as *w* in Fig. 1d of the main text) which is not apparent in the side views shown here.

The device used in this work for thermal conductivity ($k$) measurements consists of two SiNx islands. In addition, each island also incorporates an additional SiC "wing" or "side shelf", located on the side of each island opposite the SiC $k$ test sample (see Fig. 1c in the main text). The SiC wings are 120 – 300 nm thick (same as the thickness of the SiC $k$ test sample), 30 μm wide and 10 μm long, and the SiNx pads are 500 nm thick, 30 μm wide and 32 μm long. Dimensions are confirmed using scanning electron microscopy and spectroscopic ellipsometry. In addition, each island is suspended by six 300 μm-long and 2 μm-wide SiNx legs.

The key steps for fabricating the suspended device and sample are shown in Figure S1 and are summarized in detail as follows. First, a SiC film is grown on a 6-inch diameter, 675 μm thick, p-type Si wafer using low-pressure chemical vapor deposition (LPCVD), and then it is immediately annealed in a nitrogen atmosphere at the desired temperature and time. When fabricating non-annealed samples, the process should directly go to the next step after the SiC deposition. After the SiC deposition and annealing, a layer of UV210 photoresist (PR) is patterned using standard photolithography with an AMSL DUV (deep ultraviolet) stepper Model 5500/300. Subsequently, the pattern is transferred to the SiC layer by using chlorine-based reactive ion etching (RIE). At this step, the middle portion of SiC which will bridge the two islands and be used for thermal conductivity measurements shown in the figure is initially patterned to be 30 μm wide in the dimension out of the page, same as the islands, and 20 μm long. Next, a low-stress SiNx film with a thickness of 500 nm is grown via LPCVD and is patterned using the similar technique for patterning SiC, except that the etching now is performed with fluorine-based gases. Then in Step 6 the SiNx is patterned to slightly overlap and thus cover each contact edge of the SiC, with an overlap of 1 μm, for minimizing the thermal contact resistances. As such, the free central region of SiC for measuring its $k$ has length $L = 18$ μm instead of the original 20 μm, which are thus used for thermal conductivity calculations. Afterwards, a lift-off process is used to create serpentine Pt resistance thermometers (PRT) and related electrical connection lines. Specifically, the PR is first patterned using the ASML stepper, and then 5 nm of Cr and 50 nm of Pt are deposited via e-beam evaporation. The pattern is then transferred to the Cr/Pt by submerging the wafer in a lift-off solution. The following step is to fabricate the middle portion (the main test section) of the SiC to the desired width for thermal conductivity measurements. To do so a 200 nm thick $SiO_2$ hard mask is first grown via plasma-enhanced chemical vapor deposition (PECVD). The $SiO_2$ hard mask is then pattered by a Crestec CABL series ultra-high resolution electron beam nanolithography system and an STS advanced planar source oxide etch system with a high coil power in order to obtain straight sidewall profile. Such $SiO_2$ pattern will transfer to the SiC by using chlorine-based RIE. Finally, after removing the extra e-beam resist and $SiO_2$, the device is released using 24% w/v KOH solution at 80 °C for at least 2 hours.

## S2. Details of mean free paths of bulk SiC and thermal conductivity calculations

Referring to Eqs. (2) - (4) of the main text, it is important to first determine the frequency- and temperature-dependent mean free paths of high quality bulk SiC, $\Lambda_{bulk}(\omega,T)$, in order to then calculate the thermal conductivities of the polycrystalline SiC films used in this work. This intrinsic $\Lambda_{bulk}$ accounts for the phonon-phonon (also known as umklapp) scattering, $\Lambda_{umk}$, any phonon-impurity scattering (which includes scattering by the natural isotopic variability), $\Lambda_{imp}$, and a phonon-boundary scattering length scale (since any real sample is finite sized), $D$, which can be combined in parallel using Matthiessen's rule,

$$\Lambda_{bulk}^{-1} = \Lambda_{umk}^{-1} + \Lambda_{imp}^{-1} + D^{-1} \tag{S1}$$

Here, we use Rayleigh-like scattering of phonons[1, 2] to model the impurity scattering, $\Lambda_{imp}^{-1} = A_1 \omega^4 / v$, and use one common form[3] to calculate the Umklapp scattering, $\Lambda_{umk}^{-1} = B_1 \omega^2 T e^{-B_2/T} / v$. The unknown parameters $A_1$, $B_1$, $B_2$, and $D$ are determined by fitting the literature $k(T)$ data for a high-quality, macroscopic sample[4] to this kinetic theory model (see also the main text) over a wide temperature range as shown in Figure S2. For better visualizing the effects of those scattering mechanisms on the thermal conductivity of bulk SiC over the entire temperature range, we have performed a sensitivity analysis by varying each of the fitted parameters (except $B_2$ which has a similar $T$ range of sensitivity as $B_1$) by a factor of 2 as shown in Figure S2. It can be clearly seen that the parameters $B_1$ and $D$, related to umklapp scattering and boundary scattering, are the primary scattering mechanisms for the high and low temperatures, respectively. Also, $A_1$, related to impurity scattering, is important near the peak of the bulk $k(T)$ curve.

For simplicity we approximated the SiC phonon dispersion relation as isotropic and lumped the three acoustic branches into a single triply degenerate branch with an effective sound speed, $v_s$, calculated as

$$\frac{1}{v_s^2} = \frac{1}{3}\left(\frac{1}{v_{sL}^2} + \frac{2}{v_{sT}^2}\right) \tag{S2}$$

where $v_{sL}$ and $v_{sT}$ are respectively the longitudinal and transverse sound speeds obtained from the literature[5]. The shape of the dispersion is modeled by sine-type (Born-von Karman) function,

$$\omega = \omega_0 \sin\left(\frac{\pi q}{2q_0}\right) \tag{S3}$$

where $q$, $\omega_0$, and $q_0$ are the wave vector, maximum angular frequency, and cutoff wave vector, respectively. The cutoff wave vector is expressed as

$$q_0 = (6\pi^2 N)^{1/3} \tag{S4}$$

where $N$ is half of the atomic number density, corresponding to a two-atom basis. We determine $\omega_0$ by matching the small wave-vector phonon velocity with $v_s$

$$\omega_0 = \frac{2}{\pi} v_s q_0 \tag{S5}$$

Note that this model neglects the heat carried by optical phonon modes in SiC, which is a reasonable approximation since the real dispersion relation[6] is fairly flat for optical modes as compared to acoustic modes.

After fitting this model to the reference data in Fig. S2, the resulting parameters to determine $\Lambda_{bulk}(\omega,T)$ are summarized in Table S1. The model sensitivity can be assessed through the parameter-offset curves in Fig. S2, which reveal that, for the temperatures of interest in the main text, $\Lambda_{bulk}$ is dominated by phonon-phonon scattering. As $T$ decreases to 100 K, the $\Lambda_{bulk}$ model depicted in Fig. S2 also begins to show some minor contributions from impurity scattering, though this also is negligible as compared to the grain boundary scattering $\Lambda_{gb}^{-1}$ in the main text.

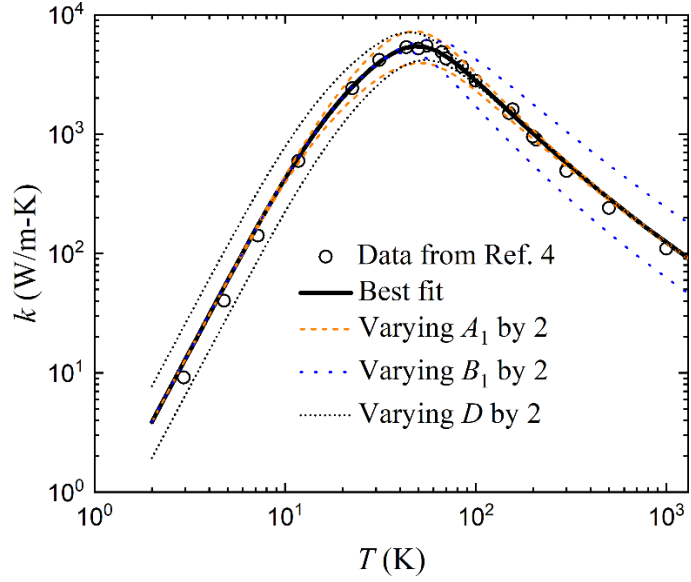

Fig. S2. The best fit model curve along with the reference data for $k(T)$ from Ref. 4. The thermal conductivity results obtained by varying each fitted parameter by a factor of 2 are also plotted for visualizing the importance of each phonon scattering mechanism.

Table S1. The calculated $\omega_0$ and $v_s$ along with the best fit parameters $A_1$, $B_1$, $B_2$, and $D$ used for thermal conductivity calculations.

| Maximum frequency, $\omega_0/2\pi$ (THz) | $v_s$ (m/s) | $A_1$ ($10^{-46}$ s$^3$) | $B_1$ ($10^{-20}$ s/K) | $B_2$ (K) | $D$ (μm) |
|---|---|---|---|---|---|
| 11.96 | 8,324 | 5.02 | 5.35 | 206 | 820 |

Table S2. The obtained average grain sizes ($D_{fit}$) from fitting the model to $k(T)$ data for all samples (#1 - #3 are repeated from Table 2 of the main text). Here the samples have been sorted by the annealing temperature, which better depicts its correlation with $D_{fit}$.

| Sample # | Thickness, $t$ (nm) | $D_{fit}$ (nm) |
|---|---|---|
| 1 (unannealed) | 300 | 5.5 ± 0.2 |
| 4 (unannealed) | 145 | 5.4 ± 0.2 |
| 8 (unannealed) | 145 | 5.4 ± 0.2 |
| 9 (unannealed) | 145 | 5.3 ± 0.2 |
| 2 (950 °C, 2 hours) | 278 | 5.6 ± 0.2 |
| 5 (950 °C, 2 hours) | 130 | 5.5 ± 0.2 |
| 3 (1100 °C, 2 hours) | 257 | 6.6 ± 0.2 |
| 6 (1100 °C, 2 hours) | 120 | 6.6 ± 0.2 |
| 7 (1100 °C, 17 hours) | 120 | 6.6 ± 0.2 |

## S3. Example of the two parameter fit ($C_1$, $C_2$)

As described in the main text the model has two free parameters, $C_1$ and $C_2$, which we determine independently for each sample by fitting its experimental $k(T)$ data. However, for all 9 samples we find in practice that $C_2$ and thus the point defect scattering is unimportant. As an example of this analysis for Sample #1, Fig. S3 shows the average residual error (defined as $\sum_{i=1}^{N} |k_{model}(T_i) - k_{measured}(T_i)|/N$, for the $N = 7$ temperatures) over a wide range of $C_1$ and $C_2$. This reveals a "confidence valley" in $C_1$-$C_2$ space. As can be clearly seen, the residual is minimized for a fairly tight range of $\frac{C_1}{2\pi}$ around $3.25 \times 10^4 \frac{m}{s}$, yet the fit loses essentially all sensitivity to the point defect scattering parameter once $\frac{C_2}{(2\pi)^4}$ exceeds around $10^{43}$ s$^{-3}$. Physically this indicates that any point defect scattering in this model is negligible compared to the other mechanisms, or equivalently, $\Lambda_{eff}^{-1} \gg \Lambda_d^{-1}$. We note that this loss of fit sensitivity for large $C_2$ is also a consequence of the impurity scattering already built into the *bulk* scattering model of Eq. (S1), since the expressions for $\Lambda_{imp}$ in Eq. (S1) and $\Lambda_d$ in Eq. (4) have the same mathematical form. Indeed, converting the $A_1$ value of Table S1 into an equivalent $C_{2,\text{bulk impurity scattering}}$ is also consistent with the "head" of the blue confidence valley seen in Fig. S3. Similar observations about the best fit being insensitive to $C_2$ are found for the confidence valleys of the other 8 samples (details omitted for brevity). Therefore, we simplify the model by dropping the defect scattering term and thus reducing the model to a one-parameter fit ($C_1$) for each sample.

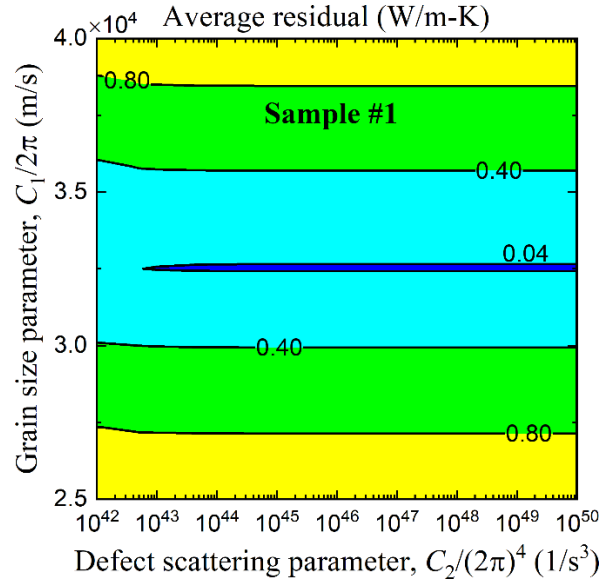

Fig. S3. Example of fit sensitivity for the two key model parameters $C_1$ and $C_2$, for Sample #1. Note that larger $C$ values correspond to longer mean free paths and thus weaker scattering. Contours are labeled by their values of the average values of the residuals comparing measured and modeled $k(T)$. As clearly seen by the dark blue "valley" of this residual surface, the smallest

residual (that is, best fit) is obtained for a specific $\frac{C_1}{2\pi}$ value around $3.25 \times 10^4 \frac{m}{s}$ but for a very wide range of $C_2$ values. This value of $C_1$ corresponds to an average grain size of $D_{fit}$ = 5.5 nm; see Eq. (5) of the main text.

**Supplementary References**